\documentstyle[aps,prc,epsfig,color,amsfonts,amssymb,twocolumn]{revtex}
\input epsf

\draft
\newcommand{\sss}{\scriptscriptstyle}

\begin{document}  
\title{Kaon photoproduction: background contributions, form factors
and missing resonances}
\author{Stijn Janssen\thanks{e-mail: stijn.janssen@rug.ac.be},
Jan Ryckebusch, Dimitri Debruyne and Tim Van Cauteren}  
\address{Department of Subatomic and Radiation Physics, \\
Ghent University, Proeftuinstraat 86, B-9000 Gent, Belgium}
\date{\today}
\maketitle
\begin{abstract} 
The photoproduction $p(\gamma, K^+)\Lambda$ process is studied within
a field-theoretic approach. It is shown that the background
contributions constitute an important part of the reaction dynamics. We
compare predictions obtained with three plausible techniques for
dealing with these background 
contributions. It appears that the extracted resonance parameters
drastically depend on the applied technique. We investigate the
implications of the corrections to the functional form of the hadronic
form factor in the contact term, recently suggested by
Davidson and Workman (Phys.~Rev.~C \textbf{63}, 025210).
The role of  background contributions and  hadronic
form factors for the identification of the quantum numbers of
``missing'' resonances is discussed.
\end{abstract}
\section{Introduction}

Gaining deeper insight into the structure of the nucleon is one of the
ultimate 
goals of current research in medium-energy physics. One of the crucial
topics is understanding the excited states of the nucleon, in brief
denoted as $N^*$.  Most of the available information concerning 
nucleon resonances is based on the knowledge extracted from
electromagnetically induced pion production and pion induced
reactions.  Despite the fact that invaluable information regarding
$N^*$'s is obtained 
in the pion sector, since long it has been realized that alternate
meson production reactions could provide additional information on the
excitation spectrum of the nucleon. In particular, the involvement of a
strange $q \overline{q}$-quark pair in the reaction opens an
additional degree of freedom and it is believed that some of the
(unobserved) resonances have specific strong coupling into these
``strange channels'' \cite{Capstick}.

At present, high-duty electron and photon facilities like CEBAF, ELSA,
MAMI, Spring8, GRAAL, Bates and LEGS provide data for electro- and
photoproduction 
of mesons with unprecedented accuracy.  One of the major challenges for
the field is extracting from the data reliable information about
resonances, like photocoupling helicity amplitudes and strong decay
widths, in an as model-independent fashion as possible.  In principle,
a complete coupled-channel analysis could handle the challenging
problem of extracting the relevant physics from the meson induced and
meson production reaction data. Over the last couple of years, there
has been 
quite some progress in this field and several frameworks to perform a
combined analysis of the photon and meson induced reactions have
been developed  \cite{Feuster2,Vrana,Manley,Chiang}. Apart from an
unified description of a wide variety of  
reactions, a coupled-channel approach  can incorporate the effect of
final state interactions in the description of the dynamics. A recent study 
\cite{Chiang} reports that these final state interaction effects on the
computed $p(\gamma,K^+)\Lambda$ cross sections are of the order of
20\%.  Apart 
from the uncertainties inherent to coupled-channel approaches, such as
unknown phase shifts and off-shell rescattering ambiguities, 
these results indicate that a large part of the $p(\gamma,K^+)\Lambda$
reaction dynamics is dominated by the first-order, 
so-called ``tree level'' diagrams. In this work, we show that in a
tree level description of the $p(\gamma,K^+)\Lambda$ process, a reliable
extraction of resonance parameters is still far from 
evident and subject to  
uncertainties.  We believe that, in addition to directing efforts
toward dealing with the coupled-channel final state interaction
effects, a proper treatment and understanding of the
first-order tree level terms is absolutely necessary. 
The primary goal of this work is to quantify the model 
dependency of the extracted resonance parameters due to the
uncertainties stemming from the background contributions and the
introduction  of hadronic form factors.  

The outline of this paper is as follows. In
Sec.~\ref{sec:reacdym}, we briefly discuss the field-theoretic
formalism to describe the reaction dynamics of the $p(\gamma, K^+)\Lambda$
process. In Sec.~\ref{sec:results} we present the results of the
numerical 
calculations. Sec.~\ref{sec:cc} provides a discussion of the influence of
background contributions. In Sec.~\ref{sec:formfac} the role of  form
factors is investigated and in Sec.~\ref{sec:misres} we address the issue
of identifying ``missing'' resonances. Our 
conclusions are presented in Sec.~\ref{sec:conclusion}.

\section{Reaction dynamics}
\label{sec:reacdym}
We describe the $p(\gamma,K^+)\Lambda$ process in
terms of hadronic degrees-of-freedom using an effective
Lagrangian approach. In this approach, every 
intermediate particle in the reaction dynamics is treated as an
effective field with its own characteristics like mass, photocoupling
amplitudes and strong decay widths.  
The effective-field theory determines the
structure of the propagators and the vertices  
which serve as input when calculating the different
Feynman graphs contributing to the reaction process. 
For the propagators of the spin 1/2 baryons, pseudo-scalar mesons and
(axial) vector mesons the standard expressions are
used. For the spin 3/2 particles, the Rarita-Schwinger form for the
propagator is adopted \cite{BenDavidson}.  For the sake of introducing
our normalization 
conventions for the coupling constants, the interaction Lagrangians
are summarized in Appendix \ref{sec:lagrangian}.
There is some ambiguity with respect to the structure of the $K
\Lambda N$-vertex in the sense that one may make use of either
pseudo-scalar or pseudo-vector 
coupling (or a combination of both). For the kaon photoproduction
process this issue has been studied by
several authors \cite{BennWright,Hsiao,Han}, but neither of the
two possible schemes has as yet been identified as favorable. In
this work, we have chosen the pseudo-scalar option.  
To account for the finite extension of the hadrons,
it is a common procedure to introduce a phenomenological form factor at
each strong vertex. These form factors depend on a cutoff mass
$\Lambda$, which sets the short-range scale of the theory. 
It is well known that the introduction of form factors breaks the
gauge invariance of the theory at the level of the Born terms and that
this can be overcome through the introduction of contact terms. Unless
specified otherwise, we have adopted the gauge restoration procedure
recently suggested by Davidson and Workman \cite{Davidson}.

In an effective-field theory, the coupling constants for each of the
individual resonances are not determined by the theory itself. They
are treated as free parameters which are extracted by performing a
global fit of the model calculations to the available data base.
In a second step, these values can be compared to quark-model
predictions, although the effect of final state interactions, which
are now absorbed 
in the effective couplings, may somehow obscure the results. 
To determine the vertex couplings, we compare our model calculations to the
SAPHIR data base 
\cite{Tran}. It contains 90 differential and 24 total cross section
points as well as 12 $\Lambda$-recoil polarization asymmetries for
photon lab energies ranging from threshold ($\omega_{lab}$ = 0.91
GeV) up to 2 GeV. When performing a
global fit to the data, the optimum set of coupling constants is the
one that produces the lowest value of $\chi^2$.  Apart from the two
main coupling constants $g_{K \Lambda p}$ and $g_{K \Sigma^0 p}$, all
the extracted resonance ($R$) parameters $G_R$ are a combination of a
photocoupling 
(sometimes called a magnetic transition moment) and a strong hadronic
coupling. A description of the various types of $G_R$ and their
connection to the Lagrangians is given in Appendix \ref{sec:lagrangian}. 

One of the striking observations when dealing with the $p(\gamma,
K^+)\Lambda$ process in terms of hadronic degrees-of-freedom, is that
the Born 
terms on their own give rise to cross sections which largely overshoot
the data.  Assuming SU(3) flavor symmetry \cite{Swart}, the coupling
constants $g_{K \Lambda p}$ and $g_{K \Sigma^0 p}$, which serve as
input parameters when computing the Born contributions, are fixed by
the well-known $g_{\pi 
NN}$ coupling.  Given the substantial mass difference between the {\em
strange} 
and the {\em up}/{\em down} quarks, it is well-known that SU(3)
symmetry is broken.  Assuming that SU(3) is broken at a 20\% level,
also the exact relation between the coupling constants is broken and
the following  ranges for the $g_{K \Lambda p}$ and $g_{K \Sigma^0 p}$
emerge: 
\begin{equation}
\begin{array}{rcccl}
-4.5\ &\leq &\ \ g_{K \Lambda p} / \sqrt{4 \pi}\ \  & \leq &\
-3.0 \;, \\ 
0.9\ & \leq &\ \ g_{K \Sigma^0 p} / \sqrt{4 \pi}\ \ & \leq
&\ 1.3 \;. 
\end{array}
\label{eq:boundcc}
\end{equation}
Using values in these ranges without any further modifications, the Born
terms inevitably produce far too much strength. This becomes clear in
Fig.~\ref{fig:born_totcs} where the computed total cross section is plotted
in a naive model that only retains point-like (this means before
introducing hadronic form factors) Born terms in the reaction process.  
Beyond doubt, the introduction of mechanisms that reduce the Born
strength is of primary 
concern to any model which aims at providing a realistic description
of the $p(\gamma, K^+)\Lambda$ process. 
In this work we present three possible schemes that accomplish this goal.
The first two schemes respect the ranges  for
the magnitude of the coupling constants imposed by (broken) SU(3)
symmetry as written in Eq.~(\ref{eq:boundcc}), the third one does not:
   
\begin{itemize}
\item The introduction of hadronic form factors is well known to
reduce the strength stemming from the Born terms.  The smaller the
cutoff mass $\Lambda$ 
the larger the reduction.  In order to sufficiently cut the
strength from the Born terms without any further modifications of the
theoretical framework, the introduction of (unrealistically) small
cutoff masses  appears necessary \cite{role_hyp,Mart2}.
\item A second option for counterbalancing the strength from the Born
terms is the introduction of hyperon resonances in the
$u$-channel \cite{AdelSagh,David}. We have shown \cite{role_hyp} that
$u$-channel hyperon resonances destructively interfere with 
the Born terms thereby reducing the total amount of strength
to a level that appears realistic.
\item A third option consists of simply ignoring the ranges for the
coupling constants of Eq.~(\ref{eq:boundcc}) \cite{Hsiao}.  This
inevitably amounts 
to using coupling constants that are significantly smaller than what is
expected on the basis of (broken) SU(3) symmetry.
\end{itemize}

Recent analyses \cite{Feuster2,Mart2,PDG} find that three 
nucleon resonances dominate the $p(\gamma,K^+)\Lambda$ reaction
dynamics: $S_{11} (1650)$, 
$P_{11}(1710)$ and $P_{13}(1720)$. The occurrence of an additional
``new'' resonance (a $D_{13}$ state) was proposed by the George
Washington group 
\cite{Mart2}.  This resonance naturally explains the structure at an
invariant mass of about 1900 MeV in the energy dependence of the total
cross section data from
the SAPHIR collaboration \cite{Tran}. The
$D_{13}(1895)$ resonance remained unobserved in pion induced and
$(\gamma, \pi)$  reactions but its existence was 
predicted by the constituent quark-model calculations of
Ref.~\cite{Capstick}.  As such, it appears as an appropriate
candidate for one of the ``missing'' resonances that have long been
sought for.

For the sake of clarity, we give here a
definition of the {\em  background} and the {\em resonant}
contributions. The term {\em resonant part} refers exclusively to the
$s$-channel (nucleon) resonance contributions. These are the $S_{11}
(1650)$, $P_{11}(1710)$, $P_{13}(1720)$ and $D_{13}(1895)$ resonances
unless specified otherwise. The Born terms, two $t$-channel
contributions involving the vector meson $K^*(892)$ and the axial
vector meson $K_1(1270)$, and
the $u$-channel hyperon resonances, which will be introduced at some
point, all contribute to what is called the {\em background}. Note
that resonances in the $t$- and $u$-channel do not ``resonate'' since
their poles are beyond the physical plane of the reaction.

\section{Results and Discussion}
\label{sec:results}

\subsection{Background contributions}
\label{sec:cc}

As detailed in Section \ref{sec:reacdym}, an effective
Lagrangian approach to the $p(\gamma, K^+)\Lambda$ process requires
additional mechanisms to counterbalance the unreasonable amounts of
strength arising from  point-like Born terms.  We have
performed model calculations using each of the three different
techniques  to deal with  the background contributions described in
the previous section. We refer to the 
three different treatments as models A, B and C and their major features
are summarized in Table~\ref{tab:chi}.

In  model A, the ``background'' is restricted to
the Born terms and $t$-channel diagrams involving the $K^*$ vector
meson and
$K_1$ axial vector meson exchange. In addition, we imposed an
under-limit of 0.4 GeV for the (freely varying) value of the
cutoff mass $\Lambda$ of the hadronic form factors used in the Born
diagrams during the fitting
procedure. It emerges that the best fits to
the data were obtained with values of $\Lambda$ that approach this
imposed under-limit corresponding with an extremely soft hadronic form
factor. 
As can be seen in Fig.~\ref{fig:totcs}(a),
the energy dependence of the background (with $\Lambda$ = 0.4 GeV)
is smooth and  steadily rising.  
Concerning the contributions from the resonant terms in model A, the
strength produced by the $P_{13}(1720)$ is rather small and the
structure about photon lab energies of 1.5 GeV is
clearly dominated by the $D_{13} (1895)$. Despite the fair agreement
with the data reached in model A, one can raise serious doubts about
the realistic character of cutoff masses as small as the kaon mass
\cite{role_hyp}. Indeed, a form factor represents a purely
phenomenological description of the short-range  dynamics and
sets a short-distance scale beyond which the theory is believed to
fail. With cutoff masses approaching the kaon mass, the 
form factor will unavoidably start playing a predominant role in the
theoretical description of the reaction dynamics, which is a rather
unsatisfactory situation for an effective theory.

In model B, we have extended the background with hyperon resonances
($\Lambda^*(1800)$ and $\Lambda^*(1810)$) in the
$u$-channel. Through destructive interference, the total background
strength gets reduced to 
acceptable levels (see Fig.~\ref{fig:totcs}(b)), a virtue which is
now reached with realistic values of the 
cutoff mass of the order 1.5 GeV.  The hyperon coupling
constants which arise from the fits are relatively large
($G_{\Lambda^*\left(1800 \right)}$ = -4.38 and $G_{\Lambda^*\left(1810
\right)}$ = -1.75) and can be subject to discussion. To clarify this
issue, we have performed fits to the data using a model which introduces
seven spin 1/2 hyperon 
resonances in the $u$-channel. The same
qualitative effect was observed but now with smaller values for
$G_{Y^*}$. In the light of these findings we argue that the two
hyperon resonances which were introduced in model B could be interpreted as
effective particles which account for a larger set of hyperon
resonances participating in the process.  
Note that $u$-channel resonances do not reach their pole and
only have a smooth energy behavior.
From Fig.~\ref{fig:totcs} it becomes clear that the 
final result for the total cross section calculated in model B displays a more
complicated pattern than what is typically observed for model A. Whereas model
A predicts that the resonances peak at their corresponding invariant
masses, in model B a rather complex interference pattern (especially at
higher photon energies) between the different resonances appears.

As a third option (model C) for controlling the magnitude of the
background contributions, we have performed a set of fits to the
data where we ignored the restrictions imposed by broken SU(3). We
only put limitations on the signs of $g_{K \Lambda p}$ and
$g_{K \Sigma^0 p}$. Completely analogous as model A, in model C the
background   
consists of the Born diagrams and the two spin 1 $t$-channel contributions. 
An under-limit of 1.1 GeV was imposed for the Born term form factor
cutoff mass  but during the fit, $\Lambda$ arrived at a rather
``hard'' value of 1.8 GeV.
Also in this model,  the data can be reasonably well described. 
Nevertheless, the 
overall best fit was obtained for a value $g_{K \Lambda p}$ =
-0.40 which is far below the SU(3) prediction of -3.75.
 
All three techniques to deal with the background terms, eventually lead
to a fair agreement of the model calculations with the available data. To 
illustrate this, Table~\ref{tab:chi} summarizes the $\chi^2$ per
degree-of-freedom obtained in the three models.  Despite the fact that
the $\chi^2$ values are 
comparable, Fig.~\ref{fig:cc_diff} clearly shows that the extracted
values for the  $N^*$ coupling constants (as defined in Appendix
\ref{sec:lagrangian}) differ drastically in the three models. 
From this observation we draw the conclusion that the model
assumptions with respect to the treatment of the background terms
heavily influence 
the extracted information about the resonances. Remarkably, it appears
that the 
choices made with respect to modeling the background terms not only
affect the magnitude of the different $N^*$ contributions, but also
the interference pattern between the overlapping resonances (see
Fig.~\ref{fig:totcs}).

In addition to the three frameworks to deal with the background
presented here,  one could think of a fourth type of model to reduce
the strength 
stemming from the Born terms: other nucleon resonances beyond the set
consisting of $S_{11}
(1650)$, $P_{11}(1710)$, $P_{13}(1720)$ and $D_{13}(1895)$ could be
introduced as likely candidates for playing a significant role in the 
$p(\gamma,K^+)\Lambda$ reaction dynamics. We have 
performed calculations introducing additional $N^*$'s in the
$s$-channel forcing the cutoff mass $\Lambda$ to adopt (realistic)
values larger than 1.1 GeV. None of the numerical calculations reached a
$\chi^2$ better than 8 (which 
has to be compared to typical values of $\chi^2 \approx$ 2.9 produced
by the other models). From these observations, we excluded this
option. In other words,
the introduction of additional resonances in the $s$-channel can not
be invoked as a viable mechanism for cutting down the background strength.

\subsection{Hadronic form factors}
\label{sec:formfac}

Due to the internal structure of the hadrons, the vertices cannot be
treated as point-like interactions. Therefor it is a widely adopted
procedure \cite{Pearce} to modify each hadronic vertex with a dipole form
factor of the type : 
\begin{equation}
F_x \left(\Lambda \right) = \frac{\Lambda^4}{\Lambda^4 + \left(x - M_x^2
\right)^2} \qquad (x \equiv s,t,u) \;.
\label{eq:formfac}
\end{equation}
Herein, $\Lambda$ is the cutoff value and $x$ represents the off-shell
momentum at the vertex.  We note that there is some arbitrariness in
the functional form of the form factor.  A major implication of
introducing hadronic form factors is that gauge invariance is broken
at the level of the Born terms.  We remark that the (axial) vector
meson and resonance exchange terms, which are 
characterized by the electromagnetic interaction Lagrangians of
the type (\ref{eq:gKVP-}), 
(\ref{eq:gKVP+}), (\ref{eq:gpr_1_2})
and (\ref{eq:gpr_3_2}), are gauge invariant by construction. As
suggested by Haberzettl \cite{Haberzettl}, 
the gauge invariance of the Born terms can be restored
by adding an additional contact term, which introduces a new form
factor $\widehat{F}$. This contact term is determined
in such a manner  
that the gauge violating terms are exactly canceled. For the
$p(\gamma, K^+)\Lambda$ case it reads:
\begin{eqnarray}
\varepsilon_\mu {\cal M}^\mu_{\mbox{\tiny contact}} &=& e g_{K \Lambda p}
\overline{u}_{\sss Y} \gamma_5 \varepsilon_\mu \left[ \frac{2 p^\mu}{s
- M^2_p} \left(\widehat{F} - F_s \right) \right. \nonumber  \\
& &\left.+ \frac{2 p_{\sss K}^\mu}{t
- M^2_{\sss K}} \left(\widehat{F} - F_t \right) \right] u_p \;,  
\label{eq:contact}
\end{eqnarray}  
where $p^\mu$ ($p_{\sss K}^\mu$) is the proton (kaon) four momentum and
$F_x(\Lambda)$ is given in Eq.~(\ref{eq:formfac}).
 
Recently, Davidson and Workman \cite{Davidson} criticized the
functional form of
$\widehat{F}$ proposed by Haberzettl (hereafter denoted by
$\widehat{F}_H$). They showed that with $\widehat{F}_H$, the contact
term of Eq.~(\ref{eq:contact}) is not 
free of poles, and consequently flawed.  At the same time, the authors
suggested an alternate recipe for the form factor (hereafter, denoted as
$\widehat{F}_{DW}$). For a detailed discussion we refer to the original
papers. Here, we just report the global form of the two different recipes:
\begin{eqnarray}
\widehat{F}_H &=& a_s F_s \left( \Lambda
\right) + a_t F_t \left( \Lambda \right) + a_u F_u \left( \Lambda
\right) \;, \label{eq:habff} \\
\widehat{F}_{DW} &=&  F_s \left( \Lambda \right) +
F_t \left( \Lambda \right) -  F_s \left( \Lambda \right) F_t \left(
\Lambda \right) \;, 
\label{eq:dwff}
\end{eqnarray}
where the $a_x$ coefficients in Eq.~(\ref{eq:habff}) have to satisfy
the relation $a_s + a_t + a_u = 1$.  

We have performed numerical calculations using both the $\widehat{F}_H$
and the $\widehat{F}_{DW}$ functional form in the contact term.
For those numerical calculations using $\widehat{F}_H$, we have put
$a_u = 0$.  This choice is 
motivated by the observation that in the $p(\gamma,K^+)\Lambda$ process,
the gauge violating terms only occur in the $s$- and $t$-channel. As a
result, calculations using the $\widehat{F}_H$ form
have two remaining free parameters ($\Lambda$ and $a_s$) stemming from
the form factors and the gauge restoring procedure.  In practice, we
found that the best fits were obtained for $a_s \approx 1$ and
accordingly 
$\widehat{F}_H \approx F_s (\Lambda)$.  
Fig.~\ref{fig:haddff} compares the values of $\widehat{F}_H$ and
$\widehat{F}_{DW}$  at various
photon energies $\omega _{lab}$ and  kaon center-of-mass
angles $\theta$. The left panels show the form factors for a cutoff
mass $\Lambda$ = 0.8 GeV, the right panels use $\Lambda$ =
1.8 GeV. They are representative for a rather ``soft'' ($\Lambda$ = 0.8 GeV)
and ``hard'' ($\Lambda$ = 1.8 GeV) option for the form factors.  It
becomes obvious 
from Fig.~\ref{fig:haddff} that the  magnitude of the form
factor $\widehat{F}$ heavily depends on the adopted recipe.

In order to assess the sensitivity of the
results to differences in the functional form of $\widehat{F}$, we
computed $p(\gamma,K^+)\Lambda$ observables using the two forms for
$\widehat{F}$.  
The results for the extracted $N^*$ coupling  
constants are given in Fig.~\ref{fig:cc_fhat} for the models
A and B.  In model A, where by construction a large role in the
reaction dynamics is 
attributed to form factors, the effect is enormous.  In model B, where
hyperon resonances are introduced to counterbalance the strength from
the Born terms and  hadronic form factors are not so dominant, the
extracted coupling constants are generally  more stable 
against variations in the functional dependence of $\widehat{F}$,
although also here sizeable variations are observed. 
A similar trend is for example seen in the photon beam asymmetry
(see Fig.~\ref{fig:phopol}). 
Whereas for model A different choices for $\widehat{F}$
even switch the sign of the predicted asymmetry, in model B the
situation looks reasonably stable. 
Only at the highest photon energies considered here, the predicted asymmetry
in model B becomes sensitive to the adopted recipe for the form factor
in the contact term. 

This conclusion is compatible with earlier observations concerning the
difference between the form factor prescriptions of Ohta 
and Haberzettl. Ohta originally suggested \cite{Ohta} to put the form factor
$\widehat{F}$ in the contact term 
equal to 1. As becomes clear from Fig.~\ref{fig:haddff}, the recipe
for $\widehat{F}$ suggested by Davidson and Workman  gives rise to values in
between those produced by the Haberzettl and Ohta form.
In several works \cite{Feuster2,Han,Haberzettl}, it was
stressed that $p(\gamma,K^+)\Lambda$ calculations with the
Haberzettl or Ohta recipe for $\widehat{F}$  can lead to very
different results. So in fact, it comes as no real surprise that the
effects are substantial.

In Ref.~\cite{Davidson_2}, Davidson and Workman studied the effect of
hadronic form factors on a multipole analysis of charged pion
production. They concluded that the extracted multipoles, for example
those listed in Ref.~\cite{Arndt}, are not heavily affected by the form
factors. Our calculations indicate that for kaon photoproduction, where
the effect of the background terms is larger than in the pion case,
great care must be taken when introducing hadronic form factors and
the corresponding gauge-restoring contact terms.

We conclude this section with a more general remark. In principle, a
correction to a hadronic form factor is not supposed to have a large
impact on the reaction dynamics. At best, hadronic form factors are a
purely phenomenological tool to smoothen the (unknown) 
high-energy behavior of the effective-field theory.  
If for some reason, the influence turns out to be large it 
is obvious that one runs into a rather unsatisfactory situation. 
In that respect, the introduction of soft hadronic form factors (model
A) in modeling the kaon photoproduction process, appears to lead to an
unacceptable level of (unphysical) model dependency in the extracted
information from fits to $p(\gamma,K^+)\Lambda$ data.

\subsection{Missing resonances}
\label{sec:misres}

The SAPHIR data \cite{Tran}, released back in 1998, made it clear that
the total $p(\gamma,K^+)\Lambda$ cross section is not characterized by
a smooth energy
dependence above the threshold peak.  The data displayed a structure
about photon lab energies of 1.5 GeV. Mart and Bennhold \cite{Mart2}
interpreted this structure as evidence for an additional
resonance and they identified it as a $D_{13}$ state with a mass of
1895 MeV. This $N^*$ state remained unobserved in pion-induced and pion
photoproduction processes but its existence and appreciable decay in
the $K^+ \Lambda$ channel was inferred from the
constituent quark calculations of Capstick and Roberts
\cite{Capstick}. As such, the $D_{13}(1895)$ appeared as a good candidate
for a ``missing'' 
resonance. Our calculations, displayed in Fig.~\ref{fig:mis_res}, 
essentially confirm the observations made in Ref.~\cite{Mart2} and
reveal that the structure at $\omega_{lab} \sim$ 1.5 GeV can be
reasonably accounted for after including in the model calculations a
$D_{13}$ resonance in the $s$-channel.  
Apart from a $D_{13}$ state, the quark-model 
calculations of Ref.\cite{Capstick} predict other $N^*$'s with  decay
in the strange 
channels in the mass range about 1.9 GeV. Other candidates 
are $S_{11}(1945)$, $P_{11}(1975)$ and $P_{13}(1950)$.  We have
performed calculations adding  a ``missing'' $P_{13}$
resonance to the basic set of $S_{11}(1650)$, $P_{11}(1710)$ and
$P_{13}(1720)$. The results of these model
calculations are also contained in Fig.~\ref{fig:mis_res}.  It is clear
that the procedure of either introducing 
an extra  $D_{13}$ or a $P_{13}$ resonance does equally well in
reproducing the 
resonant structure in the energy dependence of the total
$p(\gamma,K^+)\Lambda$ cross section, independent of the adopted
model to handle the background terms.  Similar observations were already
made in Ref.~\cite{Mart2} and there is common agreement on the fact
that the reproduction of a visual ``bump'' in the total cross section
should not be interpreted as
rock-solid evidence for the occurrence of a missing
resonance. Nevertheless, in 
Ref.\cite{Mart2} the $D_{13}$ was considered to be the
preferred candidate on the basis of the agreement between the
extracted coupling constants in the fits and the values predicted by
the quark model.  In 
the light of the discussions of the model dependencies in
Sections~\ref{sec:cc} and \ref{sec:formfac}, great care must be
exercised in drawing conclusions on the basis of the values of the
extracted coupling constants. 
Furthermore, we stress that the calculations of Mart and
Bennhold use the Haberzettl recipe for the form factor $\widehat{F}$ and
employ a relatively soft cutoff mass ($\Lambda = 0.8$ GeV) for the
Born terms. In that respect, their model comes close to what we
referred to as model A.

\section{Conclusion}
\label{sec:conclusion}

In summary, we have investigated the kaon photoproduction reaction for
photon energies up to 2~GeV in a field-theoretic approach.  From our
investigations it becomes clear that the treatment of background
processes in the $p(\gamma,K^+)\Lambda$ reaction is not free of
ambiguities.  At the same time, the background terms influence the
values of the extracted resonance parameters dramatically which makes
the extraction of model-independent information far from evident. We
have also investigated how sensitive the predictions are to the
adopted recipes for the phenomenological hadronic form factor
$\widehat{F}$ appearing in the contact terms which are meant to
restore gauge invariance.
Former investigations mainly used the Ohta or Haberzettl form for the
form factor $\widehat{F}$ .  Davidson and Workman pointed out that
both of these recipes 
are theoretically unacceptable and provided some alternate prescription.  
We have made a systematic study of the consequences of these
corrections  for the computed $p(\gamma,K^+)\Lambda$ observables. In
the energy-range under 
investigation, the corrections are rather big and the effect on the
extracted coupling constants is substantial.  Moreover, it is clear
that even an 
extensive and accurate data base as the one produced by the SAPHIR
collaboration, does not allow
to precisely determine the various contributions in the underlying
$p(\gamma,K^+)\Lambda$ reaction dynamics. The measured cross sections and
recoil polarization asymmetries do not suffice to nail down the
complicated 
interference pattern between the various resonances. After all, this
is not so surprising. It is well known that a complete meson
photoproduction experiment
needs at least seven observables to constrain the reaction amplitudes
at a fixed photon energy \cite{Baker}.  
As we find that the treatment of
background contributions and hadronic form factors heavily
influences the extracted values for the coupling constants, we deem it
premature to identify the quantum numbers of a ``missing'' resonance
on the basis of the existing data set.
Measurement of polarization observables will be
essential to further constrain the major reaction mechanisms.
 
  As a final remark, we mention that one can raise some reservations
about the applicability of 
theoretical models that handle resonances as purely individual
particles. From an historical point of view, this technique was
appropriate for pion reactions in the delta region where the process
is dominated by one resonance. In the kaon photoproduction channels,
a multitude of interfering resonances contribute and maybe one has to
develop more advanced techniques to disentangle these resonances and
their combined action.

\acknowledgments
This work was supported by the Fund for Scientific Research - Flanders
under contract number 4.0061.99.

\appendix
\section{Interaction Lagrangians}
\label{sec:lagrangian}
The interaction Lagrangians used in meson production
calculations are given in many works. For the sake of defining our
notation and normalization conventions, we summarize the ones
which are relevant for the $p(\gamma, K^+)\Lambda$ process.
\subsection{Born Terms}
The electromagnetic interaction Lagrangians for the Born terms are
given by: 
\begin{eqnarray}
{\cal L}_{\gamma pp} &=& - e \overline{N} \gamma_\mu N A^\mu
+ \frac{e \kappa_p}{4M_p}  \overline{N}  \sigma_{\mu
\nu} N F^{\mu \nu} 
\label{eq:gpp} \;,\\     
 & &\nonumber \\
{\cal L}_{\gamma \Lambda\Lambda} &=& \frac{e \kappa_{\sss
\Lambda}}{4M_{ p}} \overline{\Lambda} \sigma_{\mu \nu} \Lambda F^{\mu \nu} 
\label{eq:gyy}\;,  \\      
 & &\nonumber \\
{\cal L}_{\gamma \Lambda \Sigma^0} &=& \frac{e \kappa_{\sss
\Sigma^0 \Lambda} }{4 M_p} 
\overline{\Sigma^0}  \sigma_{\mu \nu} \Lambda F^{\mu \nu} + h.c. 
\label{eq:gyy'} \;, \\      
 & &\nonumber \\
{\cal L}_{\gamma KK} &=& - i e \left( K^\dagger \partial_\mu
K - K \partial_\mu K^\dagger \right) A^\mu \label{eq:gkk} \;.
\end{eqnarray}
The antisymmetric tensor for the photon field is defined as
$F^{\mu \nu} = \partial^\nu A^\mu - \partial^\mu A^\nu$.
For the anomalous magnetic moments we have used the values
\cite{PDG}: $\kappa_p$ = 1.793 and $\kappa_{\sss \Lambda}$ =
-0.613. For the sign of the magnetic transition moment $\kappa_{\Sigma^0
\Lambda}$, which is experimentally not accessible, we have used the de
Swart convention \cite{Swart} which yields $\kappa_{\Sigma^0 \Lambda}$
= + 1.61.
For the hadronic $K \Lambda p$ interaction, a pseudo-scalar (PS) or
pseudo-vector (PV) option is viable:
\begin{eqnarray}
{\cal L}^{PS}_{K \Lambda p} &=& -i g_{\sss K\Lambda p} K^\dagger
\overline{\Lambda} \gamma_5 N  + h.c. \;, \\
{\cal L}^{PV}_{K \Lambda p} &=& \frac{f_{K \Lambda p}}{M_{K}}
\partial^\mu K^\dagger \overline{\Lambda} \gamma_\mu \gamma_5 N +
h.c. \;.
\label{eq:kyp} 
\end{eqnarray}
All results in this work are obtained with the PS variant.

\subsection{Spin 1 Meson Exchange}

The electromagnetic coupling to a vector meson ($V$), is
described by: 
\begin{equation}
{\cal L}_{\gamma KV} = \frac{ e \kappa_{
\sss KV}}{4M} \epsilon^{\mu \nu \lambda \sigma} F_{\mu \nu}
V_{\lambda \sigma} K \;, \label{eq:gKVP-}   
\end{equation}
where the vector meson tensor is given by $V^{\mu \nu} = \partial^\nu
V^\mu - \partial^\mu V^\nu$ and $V^\mu$ is the vector field.
The photon coupling to an axial vector meson ($V_a$) reads:
\begin{equation}
{\cal L}_{\gamma KV_a}  = i \frac{ e \kappa_{\sss
KV_a}}{M} \left( \partial_\mu A_\nu \partial^\mu V_a^\nu -\partial_\mu
A_\nu \partial^\nu V_a^\mu \right) K \;, 
\label{eq:gKVP+}
\end{equation}
where $V_a^\mu$ is the axial vector field. 
The mass scale $M$ for the transition moment is arbitrary
chosen as 1.0 GeV. The complete antisymmetric tensor is defined 
as $\epsilon^{0123} = 1$. Note that this convention produces 
a sign difference to the covariant definition.
The hadronic vertex has a vector ($v$) and a tensor ($t$) part:
\begin{eqnarray}
{\cal L}_{V \Lambda p} &=& - g^v_{\sss V \Lambda p}
\overline{\Lambda} \Gamma_\mu N V^\mu  \nonumber \\
& &+ \frac{ g^t_{\sss V \Lambda p}}{2 \left(M_{\sss
\Lambda}+M_p \right)} \overline{\Lambda} \sigma_{\mu \nu} V^{\mu \nu} 
\Gamma N + h.c. \;,
\label{eq:VLP}
\end{eqnarray}
where $V$ is now a short hand notation of both a vector and an axial
vector meson. Furthermore, $\Gamma = 1 (\gamma^5)$  and
$\Gamma^\mu = \gamma^\mu$ 
$(\gamma^\mu \gamma^5)$ for vector (axial vector) meson resonances. 
The information about coupling constants which can be extracted from 
fits to the data, reads:
\begin{eqnarray}
G_V^v &=&  \frac{e g_{V \Lambda p}^v}{ 4 \pi} \kappa_{\sss KV} \;, \\ 
G_V^t &=&  \frac{e g_{V \Lambda p}^t}{ 4 \pi} \kappa_{\sss KV} \;,
\end{eqnarray}
with $V$ a vector or axial vector meson.

\subsection{Spin 1/2 Resonance Exchange}

For spin 1/2 resonances, the electromagnetic interaction reads:
\begin{equation}
{\cal L}_{\gamma BR} = \frac{e \kappa_{\sss BR} }{4 M_p}\
\overline{R} \Gamma_{\mu \nu} B + h.c. \;,
\label{eq:gpr_1_2}
\end{equation}
where the hadronic vertices are described by a pseudo-scalar (PS) or a
pseudo-vector (PV) part:
\begin{eqnarray}
{\cal L}_{K B R}^{PS} &=& - i g_{\sss K B R} K^\dagger
\overline{B} \Gamma R + h.c. \;,  \\
{\cal L}_{K B R}^{PV} &=& \frac{ f_{\sss K BR}}{M_{\sss K}}
\left( \partial^\mu K^\dagger \right) \overline{B} \Gamma_\mu R + h.c. 
\label{eq:kyp_pv} \;.    
\end{eqnarray}
Herein, $\Gamma^{\mu \nu} = \gamma^5
\sigma^{\mu \nu} (\sigma^{\mu \nu})$ for odd (even) parity
resonances. $\Gamma$ and  $\Gamma^\mu$ are defined as before. 
Further, $B$ is the baryon field (a $N$ or $\Lambda$ depending on
the corresponding vertex) and  $R$ is the spin 1/2 baryon resonance
field  (a $N^*$ or $Y^*$). In this work we have only used the PS
scheme. For spin 1/2 resonance exchange, the information regarding the
extracted coupling constant takes on the form:
\begin{equation}
G_R =  \frac{g_{\sss KBR}}{\sqrt{4 \pi}} \kappa_{\sss BR} \;.
\end{equation}

\subsection{Spin 3/2 Resonance Exchange}

For  spin 3/2 resonances, there are two terms in the Lagrangian
describing the electromagnetic  interaction:
\begin{eqnarray}
{\cal L}_{\gamma BR} &=& i \frac{e \kappa_{\sss BR}^{\left( 1
\right) }}{2M_p}\ \overline{R}^\mu \theta_{\mu \nu} \left(Y \right) 
\Gamma_\lambda B F^{\lambda \nu}\nonumber \\ 
&&-\ \frac{e \kappa_{\sss BR}^{\left( 2
\right)} }{4M_p^2}\ \overline{R}^\mu \theta_{\mu \nu} \left( X \right)
\Gamma \left( \partial_\lambda B \right) F^{\nu \lambda} + h.c. \;. 
\label{eq:gpr_3_2} 
\end{eqnarray}
The hadronic vertex is given by:
\begin{equation}
{\cal L}_{K B R} = \frac{f_{\sss K B R}}{M_{\sss K}}\
\overline{R}^\mu \theta_{\mu\nu} \left( Z \right) \Gamma' B \left 
( \partial^\nu K \right) + h.c. \;.
\end{equation}
Here, $\Gamma$ and $\Gamma^\mu$ are defined as above and $\Gamma' =
\gamma^5 (1)$ for odd (even) parity resonances.
The function $\theta_{\mu\nu} \left(V\right)$ reflects the invariance
of the free Lagrangian of a spin 3/2 field under a point
transformation \cite{BenDavidson} and is given by:
\begin{equation}
\theta_{\mu\nu} \left(V\right) = g_{\mu\nu}-\left( V+ \frac{1}{2}
\right) \gamma_\mu \gamma_\nu \;.
\end{equation}
The parameters $V = X,Y,Z$ are the so called {\em off-shell parameters}.
For spin 3/2 resonance exchange, the fits of the model calculations to
the data give access to the following combination of coupling constants:
\begin{eqnarray}
G_R^{\left( 1 \right)} &=& \frac{e f_{\sss KBR}}{4\pi}
\kappa^{\left(1 \right)}_{\sss  BR} \;, \\
G_R^{\left( 2 \right)} &=& \frac{e f_{\sss KBR}}{4\pi}
\kappa^{\left(2 \right)}_{\sss  BR} \;. 
\end{eqnarray}


\begin{onecolumn}

\begin{figure}
\centering
\epsfig{file=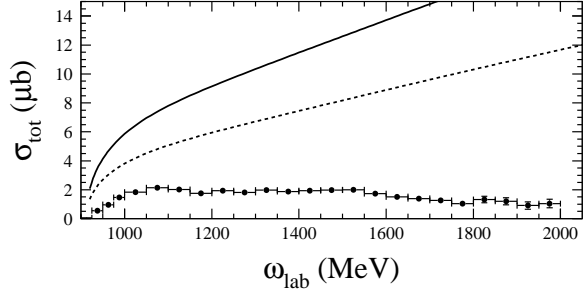,width=8cm}
\caption{The total $p(\gamma,K^+)\Lambda$  cross section as a
function of photon lab energy as obtained when solely Born terms are
included in  the reaction dynamics. No hadronic form factors were
introduced. For the solid line, the (unbroken) SU(3)
predictions for $g_{K \Lambda p}$ and $g_{K \Sigma^0 p}$ are
used while for the dashed line, the under-limit values in
Eq.~(\ref{eq:boundcc}) are taken.  The data are from
Ref.~\protect\cite{Tran}.}  
\label{fig:born_totcs}
\end{figure}

\begin{figure}
\centering
\epsfig{file=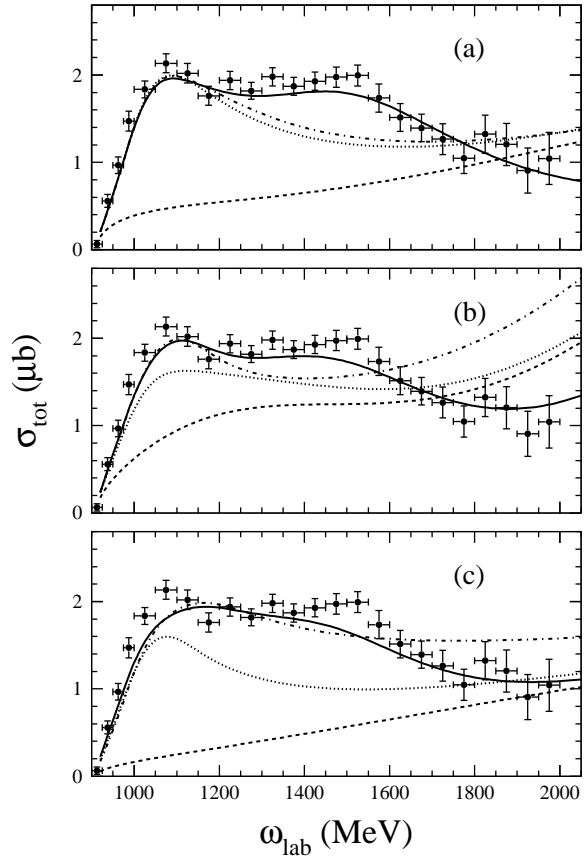,width=9cm}
\caption{The total $p(\gamma,K^+)\Lambda$  cross
section versus the photon lab energy as obtained with
three different techniques to treat the background contributions.
Panel (a), (b) and (c) use model A, B and C respectively.  In each
panel, the contribution from the background terms to the total cross
sections is denoted by the dashed line. In addition to the background
terms, the dotted line includes the $S_{11} (1650)$ and the
$P_{11}(1710)$ nucleon resonances. The dot-dashed curve adds also the
$P_{13}(1720)$ resonance. Finally, for the solid line also
the $D_{13}(1895)$ resonance is included.  The data are from Ref.~\protect
\cite{Tran}. }
\label{fig:totcs}
\end{figure}

\begin{figure}
\centering
\epsfig{file=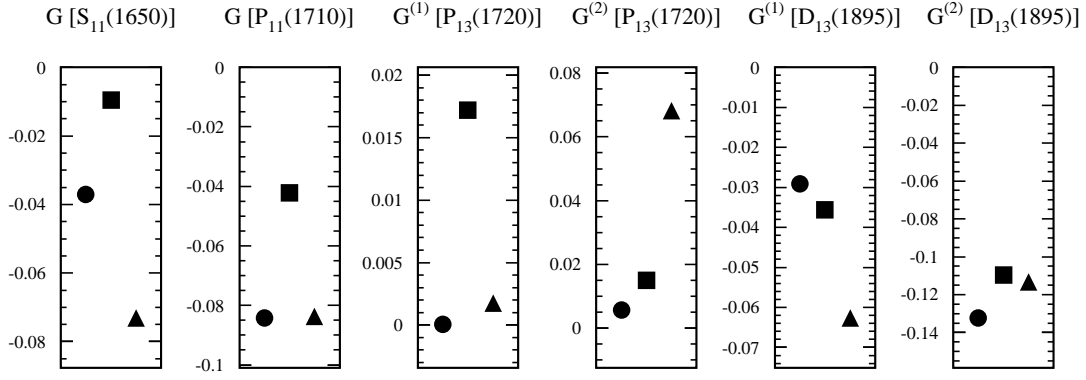,width=16cm}
\caption{The extracted coupling constants for the $S_{11}(1650)$,
$P_{11}(1710)$, $P_{13}(1720)$ and $D_{13}(1895)$ $s$-channel
resonances using three different models A, B and C for dealing with the
background terms.  The circles are for 
model A, the squares for model B and the triangles for model C.}
\label{fig:cc_diff}
\end{figure}

\begin{figure}
\centering
\epsfig{file=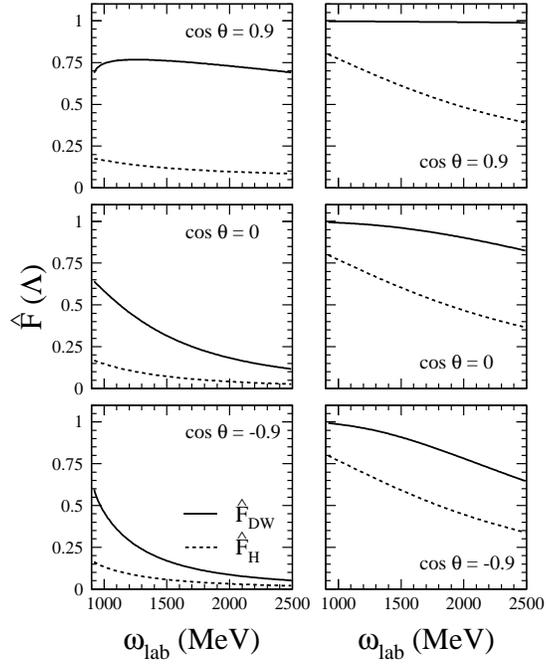,width=8cm}
\caption{The energy dependence of the hadronic form factor
$\widehat{F}$ for different kaon 
center-of-mass angles $\theta$. The left panels use $\Lambda = 0.8$
GeV and the 
right panels $\Lambda = 1.8$ GeV. The dashed line is the Haberzettl
form (with $a_s \sim 0.9$). The solid lines represent the form
proposed by Davidson and Workman.}
\label{fig:haddff}
\end{figure}

\begin{figure}
\centering
\epsfig{file=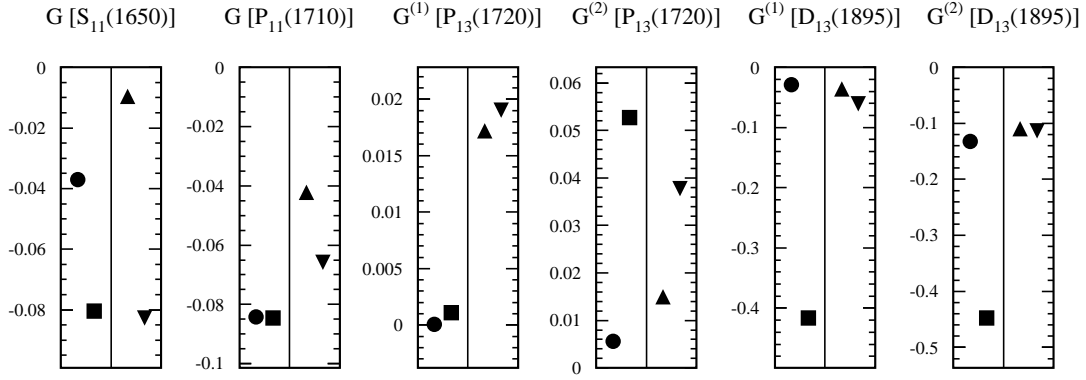,width=16cm}
\caption{The sensitivity of the extracted $N^*$ coupling constants to
the adopted form for the hadronic form 
factor $\widehat{F}$.  The circles are obtained from calculations using
the $\widehat{F}_{DW}$ form, the squares with $\widehat{F}_{H}$. 
They both correspond to a calculation which uses model A to treat the
background contributions.  Analogously, the triangles $\blacktriangle$
($\blacktriangledown$) are  
for the  $\widehat{F}_{DW}$ ($\widehat{F}_{H}$) form in model B.}
\label{fig:cc_fhat}
\end{figure}

\begin{figure}
\centering
\epsfig{file=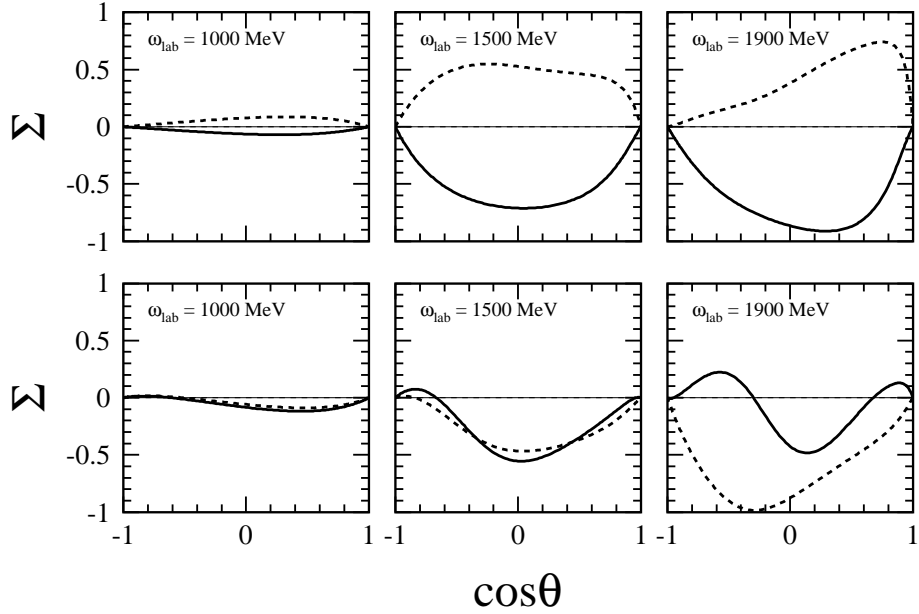,width=14cm}
\caption{The angular distribution of the beam polarization asymmetry
for $p(\vec{\gamma},K^+)\Lambda$ at three photon lab energies. The upper 
(lower) panels are results with model A (model B) for treating the
background diagrams.  Solid and dashed lines use the hadronic form
factors $\widehat{F}_{DW}$ and $\widehat{F}_H$ respectively.}
\label{fig:phopol}
\end{figure}

\begin{figure}
\centering
\epsfig{file=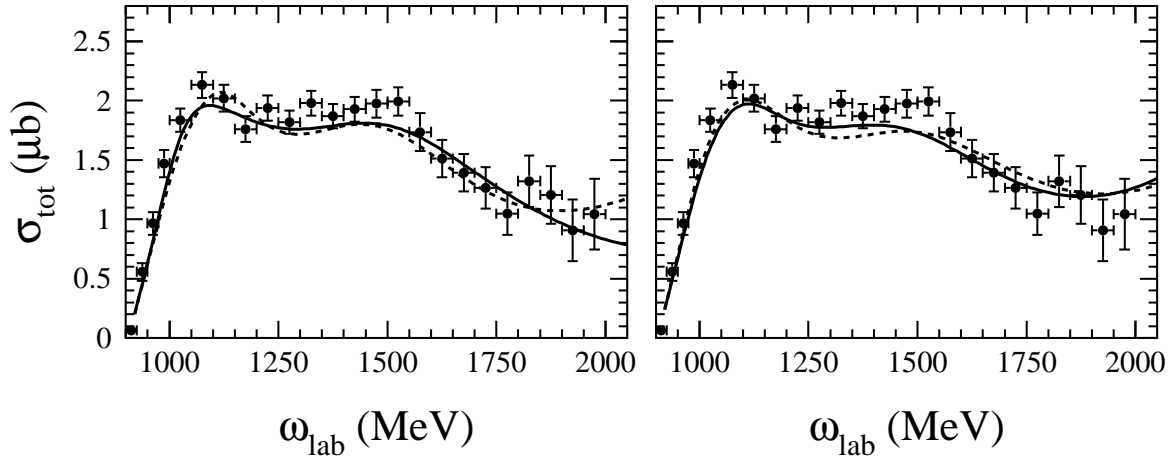,width=16cm}
\caption{Model calculations for the total $p(\gamma,K^+)\Lambda$
cross section. The 
solid curves include the ``missing'' $D_{13}$, the dashed lines include a
$P_{13}$ resonance. The left (right) panel uses model A (model B) to
describe the background contributions.}
\label{fig:mis_res}
\end{figure}

\begin{table}
\begin{center}
\begin{tabular}{cccccc}
& \\
Model & SU(3) restrictions & \multicolumn{2}{c}{$\Lambda$ cutoff mass
(GeV)} & $Y^*$ in $u$-channel & $\chi^2$ \\  
      &        & under-limit & best value  &   & \\
& \\
\hline
& \\
A & yes & $\geq$ 0.4  & 0.4 & no  &  2.99 \\
B & yes & $\geq$ 1.5  & 1.5 & $\Lambda^*(1800)$ $\Lambda^*(1810)$ &  2.89 \\
C & no  & $\geq$ 1.1  & 1.8 & no  &  2.85 \\
& \\
\end{tabular}
\caption{Schematic description of the three models used to treat the
background terms. The SU(3) restrictions for the $g_{K \Lambda p}$ 
and $g_{K \Sigma^0 p}$ coupling constants refer to the ranges
determined in Eq.~(\ref{eq:boundcc}). The cutoff masses are those of
the hadronic form factors introduced in the Born terms. 
Also given in the table is the under-limit for $\Lambda$ imposed in the
fitting procedure and  the value of $\Lambda$ corresponding with the
lowest value of $\chi^2$ (denoted as best value).
The shown $\chi^2$ values are those for the  best fit of the specific
model to  the complete SAPHIR data  set.}
\label{tab:chi}
\end{center}
\end{table}

\end{onecolumn}

\end{document}